# PHYSMATICS

Eric Zaslow[*]
Department of Mathematics
Northwestern University

## 1. Physics + Mathematics → Physmatics

Mathematics has long been used in the explanation of the physical world. It was old news even in ancient Greece:

> *The so-called Pythagoreans, who were the first to take up mathematics, not only advanced this subject, but saturated with it, they fancied that the principles of mathematics were the principles of all things.*
>
> – *Aristotle,[1]* Metaphysica

For thousands of years mathematics grew in a way that was tangible and "real." Newton made great advances in physics by incorporating and expanding upon the differential calculus of his day. Einstein's physical theories of space and time have their most natural exposition with the nineteenth century differential geometry of Poincaré. In the twentieth century, the gauge theory of particle physics and the mathematics of vector bundles grew side by side. The physicist Eugene Wigner called mathematics "unreasonably effective" in its ability to describe physics.

But developments of the twentieth century also fractured the math-physics bond. First was a mathematical tendency toward general, abstract mathematics (logic, topology, algebra, algebraic geometry). Second was the purely mathematical progress in fields which were originally tied to physics such as differential equations or geometry but were growing independently. Third was the maturation of particle theory (requiring no new mathematics) and eventual development of the so-called "standard model." In 1972, the physicist Freeman Dyson had this to say:

> *I am acutely aware of the fact that the marriage between mathematics and physics, which was so enormously fruitful in past centuries, has recently ended in divorce.*
>
> – *Freeman Dyson, Missed Opportunities, 1972*

---

[1] Biographical data for many of the people cited in this paper can be found in the appendix.

While some physicists might have disputed Dyson's bleak assessment, most now agree that the math-physics relationship has improved markedly since then. But as often happens with reconciled spouses, the nature of the relationship has undergone a shift in the process. Recent discoveries in theoretical physics shed light on the change: not only does math continue to underlie the description of physics, more and more physics is seen "underlying" pure mathematics.[2]

The new developments in mathematics and physics have been summarized by Edward Witten, the only physicist to claim the Fields Medal, the highest honor in mathematics:

> *I think there is some change. If you went back to the 19th century or earlier, mathematicians and physicists tended to be the same people. But in the 20th century, mathematics became much broader and in many ways much more abstract. What has happened in the last 20 years or so is that some areas of mathematics that seemed to be so abstract that they were no longer connected with physics instead turn out to be related to the new quantum physics, the quantum gauge theories, and especially the supersymmetric theories and string theories that physicists are developing now.*
>
> *– Edward Witten, Frontline (India), 2001*

I use the word "physmatics" to describe this new link between physics and mathematics, a link which unites the most theoretical and abstract aspects of these disciplines. This contrasts with "mathematical physics," which historically deals with concrete applications of mathematics to physics. "Mathematical physics" casts mathematics in a subordinate role. In physmatics – the word and the field – the two are equal partners. This paper aims to describe the nature of this partnership by giving an illustrative example of physmatics: duality in quantum theory and its mathematical interpretation.

## 2. Mathematics in our mist

Today as I write this, it is raining. Drops of water are forming in clouds and falling to the ground. How do they form? How do they fall? When we try to answer these questions we see the inexorable creep of mathematics into physics.

Rain formation involves the actions of zillions of small particles – droplets and the molecules therein. These particles interact with each other, coalesce, float. One cannot possibly hope to account for everything that takes place, so physicists adopt a statistical model to predict what is likely to occur. With such a large number of events, the predictions become quite accurate, on average. However, saying for certain how any given droplet will behave is an intractable problem. For instance, you may predict that a good strikeout pitcher will fan about around 200 batters in a season, but you can't know how any particular player will fare at the plate. There are lots of things in heaven and earth, so we need a good *quantitative* way of describing the average behavior of large numbers of events: it's called *statistics*.[3] Here's a picture of a recent paper in Physical Review Letters which does just that.

---

[2] Of course, physics cannot truly underlie mathematics, as mathematical definitions do not make reference to any physical structures. Perhaps physically relevant ideas are prominent in mathematics due to some cognitive attachment to reality? I couldn't say.

VOLUME 90, NUMBER 1     PHYSICAL REVIEW LETTERS     week ending 10 JANUARY 2003

**Kinetic Potential and Barrier Crossing: A Model for Warm Cloud Drizzle Formation**

Robert McGraw and Yangang Liu

*Environmental Sciences Department, Atmospheric Sciences Division, Brookhaven National Laboratory, Upton, New York 11973*
(Received 21 June 2002; published 9 January 2003)

> The kinetic potential of nucleation theory is used to describe droplet growth processes in a cloud. Drizzle formation is identified as a statistical barrier-crossing phenomenon that transforms cloud droplets to drizzle size with a rate dependent on turbulent diffusion, droplet collection, and size distribution. Steady-state and transient drizzle rates are calculated for typical cloud conditions. We find drizzle more likely under transient conditions. The model quantifies an important indirect effect of aerosols on climate-drizzle suppression in clouds of higher droplet concentration.

DOI: 10.1103/PhysRevLett.90.018501     PACS numbers: 92.60.Nv, 47.55.Dz, 82.60.Nh

Clouds and precipitation play crucial roles in regulating Earth's energy balance and water cycle [1]. Although it has been well established that three basic physical processes (nucleation, condensation, and collection) are the equilibrium:

$$A_g + A_1 = A_{g+1}, \quad (1)$$

where $A_1$ represents the water vapor monomer and $A_g$ a

Statistics helps us understand many particles, but let us focus our attention instead on the initial descent of a single water droplet. The particle speeds up (well... *down*) toward the Earth. The word "speed" suggests that we are modeling the droplet as occupying a definite location, describing that location with some coordinates of definite length, and measuring the rate of change of those coordinates with respect to some path in "time." Any model of "speeding up" would have to make use of something like these elemental parts, so at least this much mathematics authomatically creeps in. That is, we have a "coordinate system" (a way of describing positions with numbers) and a sense of distance. Further, we look for a physical rule predicting the particle's characteristics such as speed and acceleration, i.e. it's *derivatives* (speed is the distance traveled per unit time). Speed, momentum, and kinetic energy are all "geometric" in nature. The mathematics and physics needed for describing such processes are calculus and Newton's law of gravitation (Zeno, Archimedes, Euclid, Torricelli, Barrow, Kepler, Newton, Leibniz, Euler). Mathematics and physics work together to describe gravitation through geometry and calculus. These subjects are analytical, focusing on how the smallest constituents contribute to understanding.

Physics sometimes sees the big picture as well as the small, for there are quantities which don't depend on a detailed microscopic and differential treatment. Instead of rain, consider the water in a sprinkler. How would you measure how much water is coming out of your sprinkler? One way is to use a water meter: measure the water coming through the pipe. Or, if no water meter is available, collect all the water with some big tarpaulin and measure that. Both methods suffice. If you could do these things quickly, you could get a measurement of how much water comes through in a given amount of time and measure the *rate* at which water was coming out the sprinkler. In fact, in the context of electricity this rate is essentially what charge is, and how it can be measured. Electrons, protons, balls of styrofoam are like sprinklers of charge, and we can measure the amount of charge by investigating balloons which surround these objects. The total charge inside a balloon (of whatever shape) can be measured by computing the electric "flux" hitting the balloon.[4] So charge is "topological" in nature.

---

[3] Mark Twain credited the English Prime Minister Benjamin Disraeli with, "There are three kinds of lies: lies, damned lies, and statistics" – but he was probably lying. No one else has attributed the quotation to Disraeli. Twain was also wrong: statistics done right is precise.

You can look at stars, rivers, sand and snow, and begin to sense the deep relation between the physics of nature and the mathematics we need to describe it.

## 3. From mathematical physics to physmatics

In the 350 years since Newton, mathematics and physics have coevolved, now commonly recognized as distinct disciplines with some shared ground. Quantum theory, gauge theory and geometry prove to be useful terrain to explore in this context. The term "mathematical physics" has described the way in which mathematics is useful in the articulation of physical ideas. "Physical mathematics" might be used to describe the reverse process. "Physmatics" implies that the disciplines contribute equally and that the links are profound and inseparable. Here we review some of the ways mathematics and physics have worked together over the years. Our intent is not to be comprehensive, but to focus on the tools needed in subsequent sections in order to understand our main example of physmatics: duality symmetry.

Quantum Theory

Socrates believed that in our souls we know everything, and can derive knowledge by selectively and correctly "remembering." Of course, since Van Leeuwenhoek (cf. also Hooke) invented the microscope, we have known that we cannot simply infer all of what happens from observations with our naked eye. There are small things like microbes, molecules, atoms. The way these small things work demands description. If you take light from a neon lamp and pass it through a prism (which separates it into its constituent colors or "frequencies") you reveal that not all frequencies are present in the light emitted from neon atoms (see picture). In fact, the light, which is emitted when particles lose the energy they gained from the phototube, separates into distinct bands, meaning that the frequencies and energies that the particles can take appear in discrete "quanta."

---

[4] Mass is a kind of gravitational charge, and can be described similarly, by measuring the pull of gravity on a surface surrounding an object. If the object is spherical, then by symmetry you only need to measure the gravitational pull at a single point point, and Newton's law does just that.

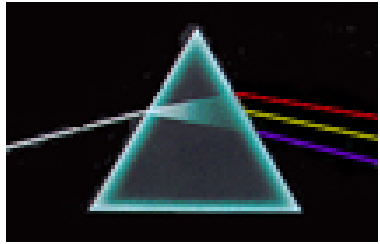

(Light from the sky is not as "pure," and contains the whole visible spectrum – so you can keep your Pink Floyd albums.) Effectively, the "speed" of the particles in their orbits gets discretized. There are other ways of observing the same phenomenon, too. Suddenly, speed does not seem to be entirely a "continuous," "differential," or "metric" quantity, when concerning "small" things like atoms.

Quantum mechanics arose as a reconciliation of the small and the large. The simple early model of de Broglie suffices for us: a particle moving around a circle can be thought of as a standing wave, so the wavelength associated to such a particle must divide the cirumference of the circle evenly. This interpretation mixes particles and waves together. As a result, only certain wavelengths are allowed. The wavelength is characterized by the integer number of times it divides the circumference. It turns out that the momentum charge is determined from the wavelength, so this means that the momentum is characterized by the same integer, and only certain discrete values are allowed. (Einstein's explanation of the photoelectric effect had supported the particle view of light. More than two centuries earlier, Newton and Hooke famously clashed on whether light was corpuscular or wavelike. Turns out they were both right!) This is true not only of any particle but any time there is a circular freedom in the system. For example, if a closed string is wound (or unwound) along a circle, its center of mass lies somewhere along the circle so represents a "circular degree of freedom," and is bound by the same rules: the center of mass momentum is thereby quantized and described by an integer.[5]

The precise position of a wiggling string is described by the excitations of the higher tones, or "harmonics," away from the center of mass. Each harmonic may be thought of as a particle, a constituent particle of the string. The center of mass is one such constituent "particle" of the string. The other harmonic resonances are also constituent particles, so a vibrating quantum string implies an infinite number of particles – a most curious consequence of string theory. To describe all these particles, quantum mechanics appeals to the mathematical disciplines of differential equations, functional analysis, and representation theory. These subjects have always had ties to physics.

In this paper, we will focus on aspects of mathematics once thought to be strictly "pure" and unrelated to physics, then find their physical manifestations. The linkage between pure math and theoretical physics lies at the heart of physmatics. What we have learned in this section is that the momentum of the center of mass of a string on a circle is described by an integer. We will use this important fact to help demonstrate physmatics. To do so, let us discuss another type of charge associated with a string on a circle. In addition to the integer describing momentum, the closed string stretched along a circle also has a "topological charge" – how many times it wraps around. More commonly, this number describes how many times your scrunchy is wound around your pony tail. This topological charge involves no quantum mechanics, and is easily seen "classically." So we see that the momentum charge and the topological charge appear on the equal footing, in that they are both described by integers (it is essential here that a circle is not infinite in extent). Yet they are quite different in nature. Our main example of duality will be a symmetry which relates these very different types of charges. For now, though, we return to the mathematics of topological charge.

---

[5] In quantum mechanics, momentum becomes quantized on a circle because the wave function must be well defined. Consider the function $\sin(ax)$, where $x$ is an angle (measured in radians) on a circle and $a$ is some number. This function represents the real part of a wave function with momentum $a$. But in order for this function to be well defined, it must have one single value at $x = 0$ and $x = 2\pi$ and $x = 4\pi$, etc. This requires that $a$ be an integer. The underlying reason is for quantization is the same as what de Broglie proposed.

## Differential Topology

Mathematics has developed way of relating "local" quantities (like derivatives) to "global" quantities (like topological charge). Appropriately enough, it is called "differential topology." Here is a taste. You can count how many laps Jeff Gordon has made in the Daytona 500 by standing at the starting line and adding one each time he passes you. Or, you can take his instantaneous speed and integrate it as a function of time (the continuous version of "rate times time equals distance") to determine the total distance; then, divide by the length of the course. The answer, if he has started and ended at the starting line, will be an integer. (The model is idealized – he doesn't weave back and forth or make pit stops here.) A more sophisticated example involves this "baaagel":

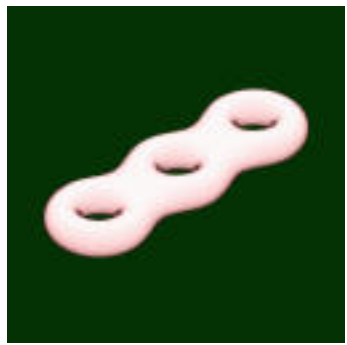

We can look at it and see there are three holes, surprisingly but we can also compute this number (3) through local calculations – and the result only depends on the number of holes, not the details of the picture. Here's how it goes. First, calculate the curvature $K$ at every point, where $K$ is the reciprocal of the product of the radii of the largest and smallest fitting circles,[6] and is negative if these circles are on opposite sides of the object. Then integrate $K$

---

[6] Imagine nailing your surface to the floor at the point in question. A fitting circle (or fitting arc) can be thought of as a coin (or portion of a coin) which fits snugly to the surface and is standing on its

over the manifold and divide by $2\pi$. The theorem of Gauss and Bonnet says that the result will be two minus twice the number of holes. Remarkably, the answer must be an integer! In this example, the answer is $2 - 2\times 3 = -4$, and does not depend on the precise shape of the baaagel, as long as it has three holes.

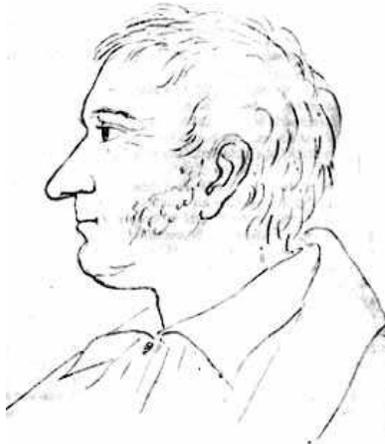

Karl Friedrich Gauss: "Few, but ripe."

For example, on a sphere, both circles have radius $R$, so we integrate $K$ and divide by $2\pi$. Since $K$ is constant on the sphere, the integral of $K$ over the surface gives $K$ times the surface area, and we get $(1/R^2)(4\pi R^2)(1/2\pi) = 2 = 2 - 2\times 0$. Indeed the sphere has no holes. In constrast, if we had an ordinary bagel, we would find a cancellation of positive and negative contributions, so the net answer is zero. $0 = 2 - 2\times 1$, and we conclude that the bagel has one hole. The actual calculation for the baaagel is somewhat more difficult.

This remarkable theorem has all sorts of generalizations – to higher dimensions, to other kinds of spaces, to invariants of other objects – many of which have importance in physics. The key feature remains the same. The input is a space and the output depends only on the topological or global nature of the space or whatever type of object is being discussed. In the case at hand, the number of holes was an important integer invariant describing the topology of a two-dimensional surface.

The lesson is that spaces can be characterized by integers which describe "global" information, and that we have concrete ways of computing them. Similar integer invariants describe the topology of the fiber bundles we will meet when we study gauge symmetry in the next section.

## Gauge Symmetry and Fiber Bundles

---

side on top of the nail head. As you rotate the coin, you need to change its size so that it may fit snugly. Call $R_1$ and $R_2$ the largest and smallest radii that you encounter over all rotations. Then $K = 1/(R_1 R_2)$. Note that by symmetry, for a sphere of radius R we will have $R_1 = R_2 = R$ at every point.

Another major theme in both mathematical physics and the physmatics we will discuss is the interplay of geometry and classical field theories. Maxwell formulated the laws of electricity and magnetism in an elegant way with beautiful equations.

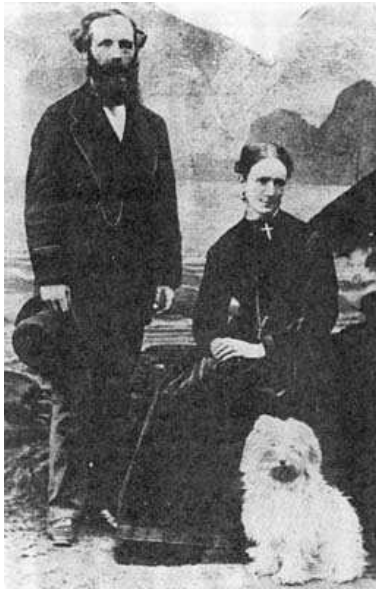
James Clerk Maxwell and wife Katherine (and dog).

In the most unified and most correct formulation, the electric field and the magnetic field are *not* fundamental. For example, a static electric field can be derived as the gradient of the electric potential (which is akin to the voltage of a circuit). So the three coordinates of the electric field vector at a point are more efficiently expressed in terms of the one value of a function (3→1). Note that we are "free" to add a constant to the potential as it does not affect the gradient, just as the overall altitude of a ski slope does not affect its steepness. The addition of a constant is an important symmetry. In fact, in electromagnetism there is a much larger freedom at play here, called "gauge symmetry." Not only can the three coordinate functions of the electric field vector be determined from a single potential function, the magnetic field is also derivable from a more fundamental magnetic potential. Together, one finds a gauge symmetry allowing a change by an arbitrary function, with no effect on physical fields. This symmetry is much greater than the shift by a constant.

Let us review how this works. Have you heard about the Suite Vollard building in Brazil? Here it is.

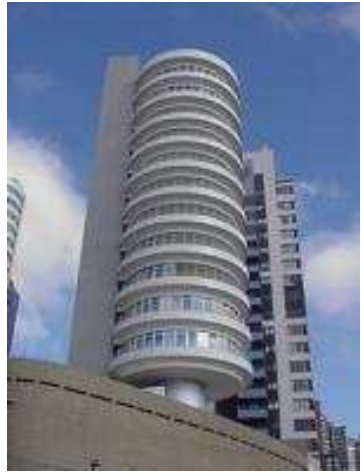

Each apartment can rotate independently by an arbitrary angle, or "phase." So the angle as a function of height is arbitrary, and we are still left with the same remarkable building. Similarly, the combined electric and magnetic potentials – $A$ – can be altered by a "gauge transformation" defined by a function which changes by an arbitrary phase at all points in spacetime. To give a toy mathematical model of independence of a function, $f$, imagine that $A = (a,b)$ from which we derived some electric quantity $E = b/a$ and a magnetic quantity $B = a/(a+b)$. If we change $A$ by $A \to (af, bf)$, then $E$ and $B$ remain the same. The mathematical picture behind this (the real model, not our toy version) is something like a giant combination lock, i.e. one circle with no natural "starting point" at every point of spacetime. Such a structure is called a "fiber bundle," the phrase being an agricultural metaphor to a sheaf of wheat: any cross section gives a slice which looks like our "base space" (the set of stalks) and is intersected once by each stalk or "fiber" (the piece of wheat). In electromagnetism, the bundle is a circle bundle, since each fiber is a circle, and the magnetic charge defined by the fields is an integer describing global information of the bundle. (The fact that the fiber is a circle is related to the quantization of electric charge, rather like the quantization of momentum.) The physicists Yang and Mills generalized this symmetry notion to include the field theories that describe particle physics, and the mathematical constructs – in which the fibers are more general than circles – have become central pillars of twentieth century mathematics.

A geometrical fiber bundle that is easy to visualize is the tangent bundle of a surface in space. The surface is the base space, and at any point of the base there is a plane of tangent vectors, the fiber.

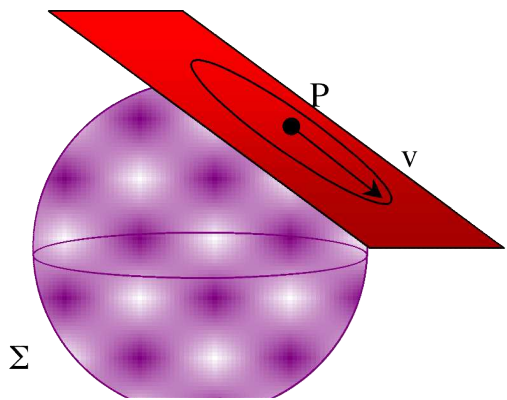

> At each point P of the surface Σ there is a whole plane (red) of tangent vectors v. This describes the tangent bundle as a vector bundle. If we consider only tangent vectors of unit length, there is a circle's worth of vectors. This describes the unit tangent bundle as a circle bundle.

Note that this plane twists around as we move along the surface. Nontrivial fiber bundles have interesting twistings, and the tangent bundle of a sphere is a good example. For another example, consider a base space which is a circle, over which we have a circle bundle. As we go around the base circle, the fiber circle can come back to itself, or it can come back rotated by a phase. The different continuous phases in this example help to describe different circle bundles which have the same integer invariants.

We have learned that the global objects relevant to gauge theory are circle bundles, and are characterized by integer-valued invariants that remain constant under perturbations – something akin to the number of holes of a baaagel. One such invariant is the physical magnetic charge. In the case of a bundle whose fibers are flat spaces and whose base space is a surface, one obvious invariant is the dimension of the fibers. Another is a more direct generalization of the number of holes of the baaagel called the *degree* of the bundle.[7] Atiyah, Patodi and Singer interpreted the integers such as degree as the number of solutions of differential equations related to the spaces involved. This interpretation gives yet another link between local and global quantities, and will be relevant to the counting of states in our example of duality.

Relativity

Physics of the very large also leads to new structures. Einstein explained the constancy of the speed of light by removing the constancy of the notion of time. The time we assign to something is a coordinate, like on a piece of graph paper. Another observer may be using graph paper which is rotated or even stretched/curved/twisted with respect to our own. That observer will measure different values of time and spatial coordinates. But let's hear it from the source:

> *Put your hand on a hot stove for a minute, and it seems like an hour. Sit with a pretty girl for an hour, and it seems like a minute. That's relativity.*
>
> *– Albert Einstein*

We should even be able to do physics if our graph paper got warped in the rain and the familiar equation $y = mx + b$ suddenly describes a *curved* line. All we should really need is a way of describing particles, in order to derive the rules which govern them. The mathematics needed for doing all this is differential geometry, the result of contributions from Riemann, Lorentz, Ricci-Curbastro, Levi-Civita and others. The mathematical idea is the same as the physical idea – if our description is sufficiently robust, we should be able to perform any geometrical calculation.

---

[7] These bundles can also be assigned continuous phases, as we shall see.

The problem with Newton's gravitation, from this point of view, is that its expression depends essentially on using a particular coordinate system. If your coordinate system is spinning, say, then a fictitious "force" arises seemingly out of nowhere. On the spinning Earth, the Coriolis force affecting water draining from a tub emerges in this way (check out http://ww2010.atmos.uiuc.edu/(Gh)/guides/mtr/fw/crls.rxml).

According to Einstein, if you release a particle into a gravitational background, it will travel in a "straight" line, where the notion of "straight" now depends on the massive bodies occupying space. So the arc of a diver is really "straight" in a sense, whereas the giant Earth has curved space itself to make straight look like a parabola – as if you had drawn a straight line on a piece of paper and then curled the paper. Likewise, when light "bends" around a star, it is really going in a "straight" line, but due to the presence of gravitational bodies what's "straight" does not look linear in our warped coordinates. We will not use Einstein's theory directly, but will make use of the generalized notion of "straight" repeatedly.

Einstein's theory of general relativity could not have been formulated without the benefit of the mathematics of differential geometry. However, that mathematics was discovered largely independently of the physics. In physmatics, there is *inter*dependence.

Physmatics

We have seen that mathematics and physics have had parallel and intermeshed histories. But nowadays we clearly distinguish these academic disciplines. We use mathematics – sometimes sophisticated and deep mathematics – to describe and quantify phenomena in physics. Mathematics has been called "unreasonably effective" in its ability to do so, but for some time there was great progress in particle physics with little or no interaction with mathematics, and this led to Dyson's "divorce."

The situation has improved markedly since then, leading to a full reconciliation – but the nature of the relationship has changed. This new relationship is what Witten was describing in the quotation from this paper's introduction. Now here is what Sir Michael Atiyah, England's foremost mathematician, has to say about Witten:

> *Although he is definitely a physicist… his command of mathematics is rivalled by few mathematicians, and his ability to interpret physical ideas in mathematical form is quite unique. Time and again he has surprised the mathematical community by his brilliant application of physical insight leading to new and deep mathematical theorems.*
>
> *– Sir Michael Atiyah*

In fact, the kinship between theoretical physics and abstract mathematics has only grown stronger – to the extent that mathematicians and physicists arrange joint conferences, routinely cross-list their papers on electronic archives, and – perhaps most tellingly – are constantly querying one another over the language and art of each other's trade.

Whereas a century ago David Hilbert declared "physics is much too hard for physicists," the situation has begun to reverse itself. Perhaps mathematics is too physical for mathematicians?

This deeper bond, this shared home, is the essence of physmatics, and is best demonstrated through the concept of "duality."

## 4. Duality in physics

String theory is a branch of physics that tries to unify quantum theory with Einstein's gravitation. Its description incorporates all the associated mathematics, and more. In fact, phenomena in string theory seem to have a kind of unifying effect in mathematics, as well. String theory is the ideal showcase for physmatics. However, we must stress that string theory remains untested in the real world, and its status as a physical theory is still undecided and a source of controversy. We conveniently ignore this controversy and focus on the physmatics of strings.[8] As for the physics of strings, we offer only this glib explanation: instead of point particles, string theory posits that the fundamental building blocks of nature are loops of string. Particle physics is recovered by considering "zero length" strings, or just focusing on the center of mass constituent particle. The wiggles account for "stringy" behavior.

We can best view the interrelationship between mathematics and physics by focusing, within the vast interplay that we have outlined, on two quantities involved in a string's motion along a circle: the center of mass momentum and the winding number. Recall that the center of mass of the string could lie anywhere on the circle; this "constituent particle" of the string has a momentum which was quantized from de Broglie's idea that the circle's circumference must equal an integer number of "wavelengths" of the particle. Likewise, a string wound around a circle has a classically quantized topological winding number that does not involve the principles of quantum mechanics. Both of these quantities involve the "nonwiggly" part of the string. That is, a string – like a taut rubber band – can be moving with momentum and winding and not look wiggly. All the wiggles (or higher harmonics) correspond to additional constituent particles, but play no essential role in what's to come – so can be ignored for our purposes. Good thing we can ignore 'em, 'cause these are The Wiggles:

---

[8] Theory preceding experiment sparks controversy in every realm. The writer Sir Arthur Conan Doyle: "It is a capital mistake to theorize before one has data." The physicist Paul Adrien Maurice Dirac who predicted antimatter: "I think that there is a moral to this story, namely that it is more important to have beauty in one's equations than to have them fit experiment." The linguist R. L. Trask on Noam Chomsky: "many critics have argued that that this retreat into ever greater abstractness is futile and self-defeating: if we make our principles sufficiently abstract and sufficiently well insulated from the observable data, then these principles become unfalsifiable and untestable. That is, any given abstract principle can be made consistent with any set of data at all, and we no longer have a testable scientific hypothesis, but only an article of faith." All these sentiments have been echoed in the debate over string theory.

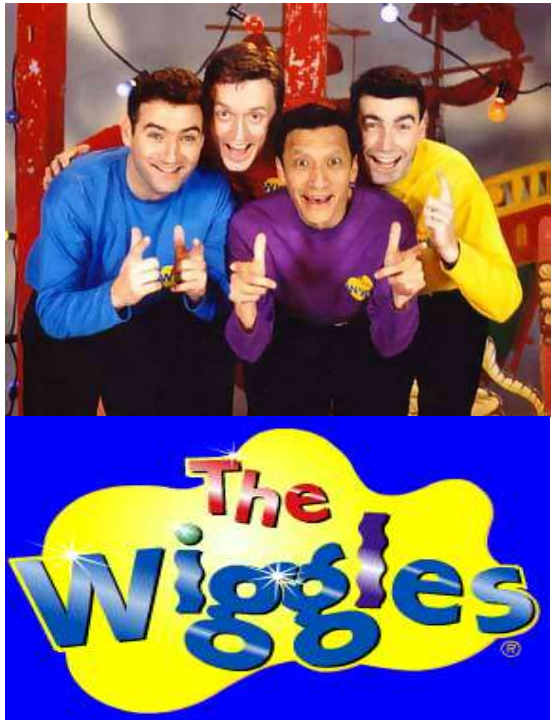

An *n*-winding string can move in such a way that, in appropriate units, its center of mass constituent "particle" has definite momentum described by the integer *m*. Such a string has an energy equal to length-squared plus momentum-squared, or $(nR)^2 + (m/R)^2$, where *R* is the radius of the circle. So by this simple equation, we see that a (3,5) string on a circle of radius *R* has the same energy as a (5,3) string on a circle of radius 1/*R*. More generally, an (*m*, *n*) string on *R* has the same energy as an (*n*, *m*) string on 1/*R*! Could it be that two circles of reciprocal radii give rise to the same "particles," after simply swapping the winding and momentum numbers of the states? Yes!! This is an important example of "duality" in physics, due to Buscher, and we will soon reap the mathematical fruits of this observation. *Duality* is when two different physical theories have the same structure, and the example of two circles of inverse radii is the most basic in physmatics. Before continuing with this example, we will first explore the notion of duality in some simpler contexts.

Duality in TV: "The Honeymooners" has the same structure as "The Flintstones." Not only is there a correspondence among principal characters, the way in which these characters relate to one another is the same (blustery but good-natured husband, level-headed and tolerant wife, overly eager neighbor with bland wife). "The Mary Tyler Moore Show" has nearly the same structure as "Murphy Brown." (Cf. also "The Wonder Years" and "Oliver Beane," "Married... With Children" and "Unhappily Ever After.")

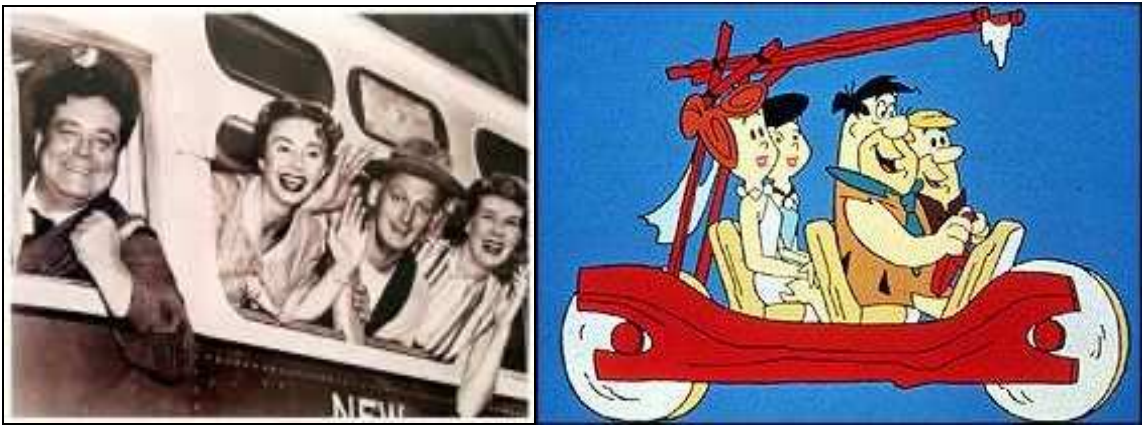

Duality in everyday math: $10^2 \times 10^3 = 10^5$. This statement is "dual" to $2 + 3 = 5$. Under the correspondence, a positive number is mapped to its logarithm. The operation of multiplication is mapped to addition. The relevant structures – multiplication and addition – are preserved: $ab=c$ if and only if $\log(a) + \log(b) = \log(c)$.

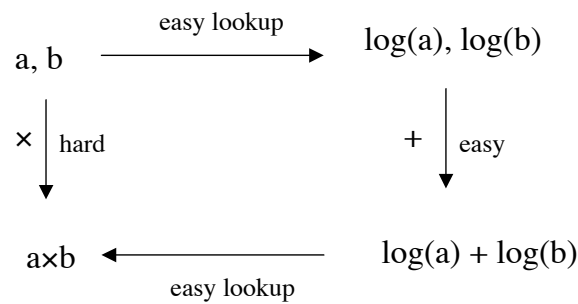

In the days of slide rules, we used this duality because addition is easier than multiplication. Map the problem to something easier, then map your answer back.

We are now able to discuss the consequence of duality in physics involving strings on circles of inverse radii. Imagine a world in which we live on a giant circle.[9] In such a world, physical particles made up of pure momentum states (with winding number zero) would have very little energy. This follows from our equation: when winding $n = 0$, we have $E = (m/R)^2$, which is small if $R$ is very large. Such states could be easy to assemble. Using little energy, we could construct "packets" of such states that look like ordinary particles, objects. Physical states involving non-zero winding numbers would have extremely large energy. We could effectively ignore these high-energy states, since they would be way out of our range – just as a ten-foot ceiling would have little effect on the lives of Lilliputians. We might then build a particle $X$ that has no total winding number through some physical process at ordinary energies, and measure it to have total *momentum* numbered by *m*. Now by the duality found, there corresponds a *dual* world on a dual *tiny* circle. The particle dual to X will have no total momentum, as momentum and winding get switched. In this dual world the low-energy states are the dual ones, i.e. winding states with zero momentum. Any non-zero momentum would have enormous energy, as $(n/R)^2$ is now large. In the dual world, our dual motions would create the particle *dual* to *X* and we will measure it to have *winding* number *m*. The map of particle states is easy: (*a*,*b*) is dual to (*b*,*a*). The point is that all the dual calculations in the tiny circle world would be exactly equal to the calculations in the large circle world. We would have no way of determining which world we lived in.

---

[9] If you do not think yourself thin enough to fit on a one-dimensional circle, you may think of the Cartesian product of three circles, which looks just like ordinary three-dimensional space except that if you walked too far in a coordinate direction, you come back to where you started (as in some video games). Similarly, the surface of a bagel is the product of two circles.

More intriguing examples abound. Sometimes, the dual spaces don't even look remotely alike: they differ in *topology* as well as size and shape – as different as bagel and bialy. The correspondence between the states is not always so easy, either. "Mirror symmetry," due to Buscher, Lykken, Greene, Plesser, Vafa and others, involves dualities of this type. The "mirror" in the name represents something akin to the reversal $(a,b) \leftrightarrow (b,a)$.

[We note there are even more radical types of duality, in which different *types* of physical theories turn out to be equivalent. Examples are heterotic-TypeII duality and the AdS/CFT correspondence. The AdS/CFT correspondence, for example, states that a particular type of string theory is actually equivalent to a specific gauge theory *without* strings!]

A duality of a robust model of physics demands a correspondence among a vast array of substructures. For example, our physical world has charged particles, black holes, and even the possibility of magnetically charged "monopoles." If there were a dual theory that had the same look, would these objects be mixed up under the duality map? Would black holes be mapped to charged particles? We can't just map particles willy-nilly. We have to preserve the relationships, as well! (So not every TV show with six characters would be considered equivalent just because we could map the characters to each other -- though admittedly many such shows are equivalently vapid.)

We conclude that a duality is a correspondence of objects and an equivalence of the way in which the objects relate. This way of phrasing it makes the mathematical concept clear: equivalence of categories. After learning the mathematics, we will apply it to the product of two circles: a bagel.

## 5. Categories in mathematics

*No one really understood music unless he was a scientist, her father had declared, and not just a scientist, either, oh, no, only the real ones, the theoreticians, whose language mathematics. She had not understood mathematics until he had explained to her that it was the symbolic language of relationships. "And relationships," he had told her, "contained the essential meaning of life."*

    *– Pearl Buck, The Goddess Abides, Pt. I, 1972.*

*Mathematicians do not study objects, but relations between objects. Thus, they are free to replace some objects by others so long as the relations remain unchanged. Content to them is irrelevant: they are interested in form only.*

*– Jules Henri Poincaré*

Categories are part of a most abstract piece of mathematics, and it is amusing and enlightening to have them relate to physical concepts. You can think of a category as a framework for understanding structure – a template. We use templates in real life. Every play has its cast of characters and their relationships. Every book has the same. A sports team, too, plus some strategies. A language has words, its grammar encodes how they interact. A boat has its objects – jib, winch, tiller, hull, boom, lines, rutter – and the objects relate to one another: jerking the tiller toward you when sailing downwind will cause the boom to whip across, so duck! Decorating. Marketing. Sewing. Insurance. Architecture. Every subject has a wealth of players/characters/objects and many complicated relationships among them. We relate two subjects through analogy and metaphor: summer days and bowls of cherries bear little superficial relationship to women and life, but poets have profited by likening their structures. Mathematical sheaves are non-nutritious, but the terminology creates a useful comparison.

In mathematics, the framework of a category is extremely simplified, yet still rich enough for most purposes. Technically, a category is a bunch of objects and "maps" between them (the "maps" do not have to be actual maps, which is why we have the quotation marks). A "map" from A to B can be combined with a "map" from B to C to get a "map" from A to C. Sounds reasonable. Here are some examples.

1. The category of Sets, whose objects are sets and whose maps are maps of sets. A map $f$ from set A to set B assigns to each $x$ in A a unique element $f(x)$ lying in B. For example, $\{a,b\}$ and $\{c,d,e\}$ are sets, and the assignment ($a{\to}e, b{\to}c$), or in other words $f(a)=e$ and $f(b)=c$, defines a map $f$ going from $\{a,b\}$ to $\{c,d,e\}$. A map $f$ from A to B can be combined with a map $g$ from B to C to get a map from A to C defined by the rule $x{\to}g(f(x))$.

2. The category of topological spaces, whose objects are topological spaces, and whose maps are continuous maps of spaces. (You can think of topological spaces as sets where the elements have a notion of nearness.)

3. The category of groups, whose objects are groups and whose maps are maps of group elements which respect the product structure. (Groups are sets of *operations*, such as right-angle rotations of a square or the 3-d rotations of a sphere, or even numerical operations.) The map from the multiplicative group of positive real numbers to the additve group of all real numbers, defined by the logarithm, is a legitimate "map" in the category of groups.

(In all of these examples, the "maps" are actual maps. Not so later on.)

Both topological spaces and groups are actually sets with other structures (topology, group product). Therefore, we can construct mappings from these categories ("functors") to the category of sets, whereby we simply forget the additional structures. These mappings do not define equivalent categories, since there may be several groups associated to a given set.

For example, the set $S = \{a,b,c,d\}$ admits these two possible multiplication tables, each of which obeys the notion of multiplicative associativity (i.e., $(a \times b) \times c = a \times (b \times c)$, etc.):

*A*

| × | *a* | *b* | *c* | *d* |
|---|---|---|---|---|
| *a* | *a* | *b* | *c* | *d* |
| *b* | *b* | *a* | *d* | *c* |
| *c* | *c* | *d* | *a* | *b* |
| *d* | *d* | *c* | *b* | *a* |

*B*

| × | *a* | *b* | *c* | *d* |
|---|---|---|---|---|
| *a* | *a* | *b* | *c* | *d* |
| *b* | *b* | *c* | *d* | *a* |
| *c* | *c* | *d* | *a* | *b* |
| *d* | *d* | *a* | *b* | *c* |

(There are other possibilities, but they wind up corresponding to simple relabelings of the letters.) When we "forget" the group structure, the *different* groups *A* and *B* get mapped to the *same* set, *S*. So *S* cannot correspond to a single group, and the categories are not equivalent.

Here is an example of two categories which *are* equivalent. Fix a set *S* (any set, but you can consider the set of integers from *1* to *10*, for example). We define the category *A* whose objects are subsets of *S* and whose "maps" are inclusions of subsets – so if the subset *a* is not contained in the subset *b*, there are no "maps:" if it is, there is a single map of inclusion. For example, $\{1,2,5\}$ has a unique map to $\{1,2,4,5,8\}$, where the *1* inside the smaller set is mapped to the *1* inside the larger set, and so on. Now let's define another category *B* whose objects are functions which take values *0* or *1*. If one function f is less than or equal to another function g for every element of the set ($f \leq g$, that is $f(x) \leq g(x)$ for all *x* in *S*) then we say there is a unique "map" from *f* to *g*. Otherwise we say there are no maps between the objects *f* and *g*. The correspondence associates to a function *f* the subset *a* of points on which *f* takes value *1*, that is, $a = \{x : f(x) = 1\}$. If *f* and *g* are two functions, write *b* for the set of points where $g = 1$. Then $f \leq g$ precisely when *a* is a subset of *b*.

This is interesting, but it's becoming… um… a tad "abstract," shall we say? Surprisingly, all this formalism has a concrete home in physics.

## 6. Duality, categories and mirror symmetry

There is an important physical category that appears in the dual theories we will discuss – the category of "branes."  In certain string theories, we can construct a category of branes whose objects are the special locations where strings can end. Suppose the strings are allowed to move in a space S.  The branes will correspond to subspaces L inside S that have special properties (depending on the precise theory being considered). The "maps" between brane objects correspond to minimal strings stretching between the objects.  Strings between $A \to B$ and between $B \to C$ can join together in physical processes (tip merging with tail) to make one (or more) strings between $A \to C$. So while the "maps" are not really maps of sets, they can be combined in the required way to define a legitimate category.

Here's the kicker.  All those objects we discussed – black holes, monopoles, particles with charge – can be considered as branes, in the appropriate context.  Roughly speaking, the objects are places where there is a concentration of energy.  The strings may end on these special locations, which means they are branes.  One important way that a brane is identified as a more familiar type of object is in fact by checking that it has the same charges, i.e. appears on the same topological footing as known objects.  We also note that a quantum theory of gravity should, after all, treat black holes on a parallel footing as elementary particles:  for if you have a sufficiently massive particle, it may create a gravitational distortion akin to a black-hole.  The duality symmetry known as mirror symmetry is an equivalence of string theories associated to two different spaces.  Each of these string theories has an associated category of branes.  Mirror symmetry, then, becomes the statement that *the categories of branes in the two string theories on dual spaces are equivalent!*  This description of mirror symmetry is due to the mathematician Kontsevich and, from the physical point of view, Ferrara, Strominger, and Vafa.

Here's an important and nontrivial example of brane correspondence.  Consider the surface of a bagel.  We may think of this surface as the product of two circles (see picture), since a point can be expressed by giving a pair of angles – one along the vertical circle and one along the horizontal.

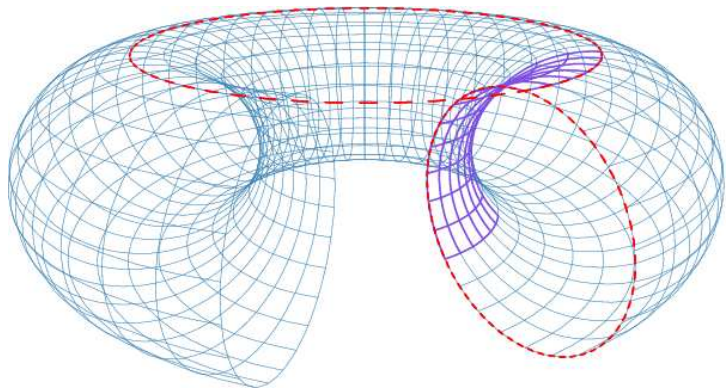

Let us now assign fixed radii to each of these two circles.[10]  We can now use our circle duality to invert the radius of one of the circles: R→1/R. The dual spaces are both bagels ("tori") but have different sizes and shapes. (In more complicated examples, the dual spaces are genuinely different, even topologically; as different as bagel and bialy.) Now since the string theories are equivalent, the branes in these spaces must have dual descriptions, an in fact the two categories of branes must be equivalent. We will demonstrate a complete correspondence among the branes, i.e. the objects. We will show that each brane has a dual partner in the dual theory, and the way it moves has a dual description as well. Further, we will see evidence that the relationships among dual pairs of branes are the same. In fact, in this example one can give a complete mathematical proof of the equivalence of brane categories, including the combinations of "maps," though it is beyond the scope of this paper.

---

[10] In the picture, only the vertical circle has a fixed radius.  The different horizontal slices have different radii.  Unfortunately, one cannot draw a picture of the particular bagel we are using in our example, but one can visualize it as a flat rectangle where the top and bottom are considered the same line (this gluing would make a cylinder) and then the left and right sides are also identified.

On bagel *A*, we study a theory in which branes have odd dimension. Since the branes are subspaces of a two-dimensional space, this means they are one-dimensional. On bagel *B*, in the dual theory, the branes have even dimension. Therefore, they are zero-dimensional points or fill the whole two-dimensions.[11] But wait! There's more! At the end of the string, where the brane is, there may be a charge. We now know that charge is associated to a circle bundle. The circle bundles then lie over the branes.

It is most productive to consider minimal energy branes. In the one-dimensional (*A*) case, this means that the lines should be "straight," which means they wrap the two circles at constant rates (or straight in usual sense, using the rectangle model of the bagel described in the footnote). The "straight" line must form a closed loop on the torus. The loop can be classified by following how many times it winds around the vertical circle (say *d*) and how many times it winds around the horizontal circle (say *r*) before closing up. The data (*r*,*d*) does not specify the loop, however, since you can move it along the surface. Thus we require an additional angle $\theta_1$ to specify the loop (sliding it along its own direction does nothing, so there is one, not two, ways to get different loops). The additional circle bundle data means that we should specify an additional angle $\theta_2$ specifying the rotation of the fiber circle when gluing both "ends" of the loop together. Here is a picture of an (*r*,*d*) = (1,17) loop:

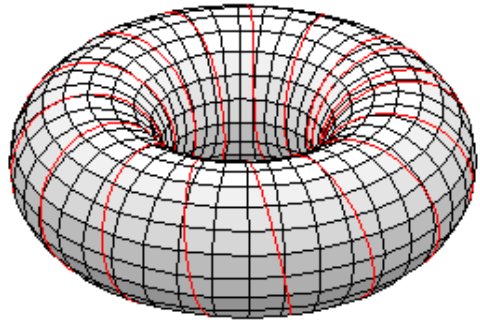

On the *B* bagel, even-dimensional subspaces can have zero or two dimensions. The two-dimensional closed spaces fill the whole bagel, but there is the extra data from the charges.

---

[11] The original models of string theory were designed to explain how charged quarks were stuck together by "tubes" with the charges at the ends. These "flux tube" models were since abandoned in favor of strings as the fundamental things.

There is the discrete, topological bundle data described in our exposition of differential topology: the dimension, or "rank," of the fibers ($r$), and the degree ($d$) of the bundle, which is an integer invariant generalizing the number of holes.[12] Atiyah proved that when $r$ and $d$ are reduced to have no common factors ("relatively prime"), there is an essentially unique bundle. (Note that the ($r,d$) from the branes of bagel A will likewise have no common factors, since that would correspond to traversing the loop more than once.) The extra data needed to specify the bundle uniquely are two phases ($\theta_1$, $\theta_2$) of a circle bundle – *two* phases, since the brane occupies the whole bagel, a product of *two* circles. Finally, the zero-dimensional branes have no fiber over the general point ("$r = 0$"), but there are now two angular variables associated to the position of the point.

We learn that branes in both models are described by a pair of relatively prime numbers ($r,d$) and have precisely two angular parameters ($\theta_1$, $\theta_2$). This tells us how to make the correspondence of objects in the brane categories precise: a loop with windings ($r,d$) is mapped to a bundle of rank $r$ and degree $d$. What about the relationships between objects, the "maps"? An open string going from one brane to another serves as a map between branes, so is a measure of their relationship. We will therefore check the correspondence of relationships by counting minimal open strings between corresponding pairs of branes. The number should be the same for dual pairs. The minimal open strings are the strings of zero length which lie at the intersection points of branes. These strings of zero length have no wiggles involving higher harmonics, and can be identified with their center-of-mass constituent point particles.

Let us begin with the *B* bagel and consider a pair of two-dimensional branes described by bundles ($r,d$) and ($r',d'$) with whatever angular phases. Since the branes fill the whole space, the zero-length open strings can lie anywhere on the bagel *B*. The minimal states are described by the quantum mechanics of the center-of-mass point particles, and the formula which counts them is one of the important generalizations of the Gauss-Bonnet theorem. The formula states that the number of minimal strings is equal to $|rd' - r'd|$. Now we check the correspondence of relationships by going to bagel *A*, where we consider one-dimensional branes with windings ($r,d$) and ($r',d'$). The minimal open strings have zero length and lie at brane intersections. It is an exercise in geometry (simple plane geometry, if you use the rectangle model of the torus) to compute that "straight" loops with windings ($r,d$) and ($r',d'$) intersect at precisely $|rd' - r'd|$ points!

Therefore, the relationships among branes are preserved under the mapping of objects. This leads us closer to concluding that the seemingly very different categories of branes on bagels *A* and *B* – one involving odd dimensional spaces, the other even – are indeed equivalent. To go further, one must check that corresponding "maps" combine together correspondingly, as in the example of the slide rule. In fact, it all works out, and Kontsevich's view of mirror symmetry as a correspondence of brane categories has been realized. We have therefore witnessed first hand a very nontrivial example of physmatics in action.

---

[12] The degree captures how much the bundle twists. For a circle bundle over a sphere, you can visualize it as follows. Consider a trivial circle bundle over the northern hemisphere, namely the product of the northern hemisphere with a circle (so the fibers are described by points of the *same* circle). Now consider the same thing for the southern hemisphere. We can obtain a circle bundle over the whole space by gluing at the equator and identifying the fibers. However, instead of just gluing the two fiber circles together along the equator, we can twist them relative to each other. As we move around the equator, the twisting angle can traverse any number of full cycles, just like the scrunchy. That integer is the degree $d$.

Such a correspondence – and its generalizations to higher-dimensional spaces, including mirror pairs of spaces which are topologically distinct – was completely unanticipated by mathematicians. The input from physics is crucial. But the framework, the definitions of the categories, and many of the methods of topological field theories which characterize mirror symmetry come from mathematics. Nowadays, scientists from both camps continue to try to understand the profound link that mirror symmetry implies in physmatics.

## 7.            A-infinity… and beyond!

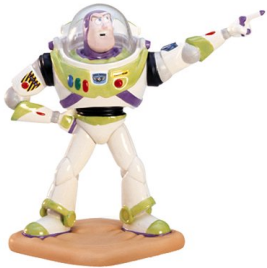

Congratulations! We've just learned the gist of mirror symmetry. But although we've come a long way already, we're actually in the middle of a marathon! I won't make you finish it, but it is important to recognize the full length of the course.

Right now in the field advances come from physicists[13] and mathematicians,[14] as well as some who fall in-between. Here is an example of some further directions. When you, your sister and her best friend from college get together, a special vibe kicks in, a group dynamic that just doesn't occur when any two of you are alone. That night at The Kilbirnie tells the story. The chemistry among the three of you is something special (how else can you explain the midnight dance of the nachos?) and not derivable in terms of pairwise relationships. The Borromean rings below are a classic example of trivial pairwise relationships (no two rings are linked) but a nontrivial three-way grouping, as you can't pull them apart (picture from http://www.popmath.uk/scupmath/pagesm/borings.html):

---

[13] E.g., Dijkgraaf, Hori, Kapustin, Klemm, Nekrasov, Moore, Vafa, Verlinde, Witten, ….
[14] E.g., Atiyah, Donaldson, Fukaya, Gross, Kontsevich, Orlov, Pandharipande, Seidel, Yau….

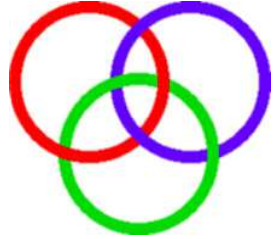

There is no room for such collective behavior in the simplified framework of a category, but there is a generalization of this framework that allows for three-fold, four-fold and even $n$-fold relationships for all $n$. Just as the pairwise relationship in "The Honeymooners" gave us an intuitive grasp of categories, $n$-fold group dynamics give a parallel notion for the so-called "A-infinity categories." These extensions of categories are relevant in mirror symmetry, and their consequences are being hotly investigated.

In mirror symmetry, duality symmetry, and in many other recent discoveries, physmatics continues unabated. And those who may scoff at the abstraction—

> *Now I feel as if I should succeed in doing something in mathematics, although I cannot see why it is so very important... The knowledge doesn't make life any sweeter or happier, does it?*
>
> —Helen Keller, The Story of My Life, 1903.

might consider this "1000-word" retort from Sir Michael:

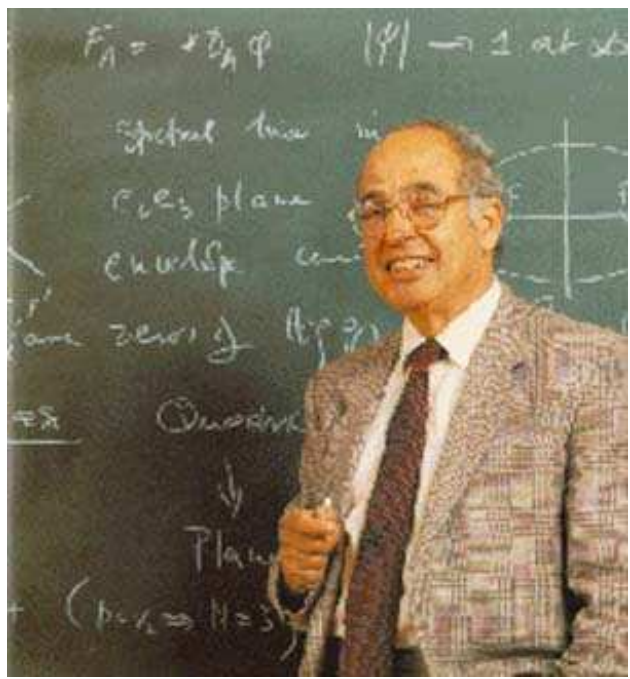

The End

**Appendix: Selected Biographical Data**

(Country indicates place of birth, not necessarily citizenship.)

Aristotle (Greece, 384–322 BC)
Atiyah, Sir Michael (England, 1929—)
Bohr, Niels (Denmark, 1885–1962)
Bonnet, Pierre Ossian (France, 1819–1892)
de Broglie, Louis (France, 1892–1987)
Buck, Pearl (USA, 1892–1973)
Chomsky, Noam (USA, 1928—)
Disraeli, Benjamin (England, )
Doyle, Sir Arthur Conan (England, 1859–1930)
Dirac, Paul Adrien Maurice (England, 1902–1984)
Disraeli, Benjamin (1804–1881)
Dyson, Freeman (England, 1923—)
Euler, Leonhard (Switzerland, 1707–1783)
Gauss, Karl Friedrich (Germany, 1777–1855)
Heisenberg, Werner (Germany, 1901–1976)
Hilbert, David (Germany/Prussia, 1862–1943)
Hooke, Robert (England, 1635–1703)
Keller, Helen (USA, 1880–1968)
Kontsevich, Maxim (Russia, 1964—)
van Leeuwenhoek, Antony (Holland, 1632–1723)
Leibnitz, Gottfried Wilhelm (Germany, 1646–1716)
Levi-Civita, Tullio (Italy, 1873–1941)
Lorentz, Hendrik Antoon (Holland, 1853–1928)
Patodi, Vijay Kumar (India, 1945–1976)
Maxwell, James Clerk (Scotland, 1831–1879)
Newton, Sir Isaac (England, 1643–1727)
Planck, Max Karl Ernst Ludwig (Germany, 1858–1947)
Poincaré, Jules Henri (France, 1854–1912)
Ricci-Curbastro, Gregorio (Italy, 1853–1925)
Riemann, Georg Friedrich Bernhard (Germany, 1826–1866)
Singer, Isadore (USA, 1924—)
Socrates (Greece, c. 470–400 BC)
Twain, Mark (Clemens, Samuel Langhorne) (USA, 1835–1910)
Vafa, Cumrun (Iran, 1960—)
Wigner, Eugene (Hungary, 1902–1995)
Witten, Edward (USA, 1951—)